\documentclass[twocolumn,aip,showpacs,preprintnumbers,amsmath,amssymb,superscriptaddress,reprint]{revtex4-1}[11pt]

\usepackage[pdftex]{graphicx}
\usepackage{amsmath}
\usepackage{amsfonts}
\usepackage{amssymb}
\usepackage{color}
\usepackage{float}
\usepackage{color}
\begin{document}
\title{Enhancement of Spin-transfer torque switching via resonant tunneling}
\author{Niladri Chatterji}
\affiliation{Department of Physics, Indian Institute of Technology Bombay, Powai, Mumbai-400076, India}
\author{Ashwin. A. Tulapurkar}
\author{Bhaskaran Muralidharan}
\affiliation{Center of Excellence in Nanoelectronics, Department of Electrical Engineering, Indian Institute of Technology Bombay, Powai, Mumbai-400076, India}
\date{\today}
\medskip
\widetext
\begin{abstract}
We propose the use of resonant tunneling as a route to enhance the spin-transfer torque switching characteristics of magnetic tunnel junctions. The proposed device structure is a resonant tunneling magnetic tunnel junction based on a MgO-semiconductor heterostructure sandwiched between a fixed magnet and a free magnet. Using the non-equilibrium Green's function formalism coupled self consistently with the Landau-Lifshitz-Gilbert-Slonczewski equation, we demonstrate enhanced tunnel magneto-resistance characteristics as well as lower switching voltages in comparison with traditional trilayer devices. Two device designs based on MgO based heterostructures are presented, where the physics of resonant tunneling leads to an enhanced spin transfer torque thereby reducing the critical switching voltage by up to 44\%. It is envisioned that the proof-of-concept presented here may lead to practical device designs via rigorous materials and interface studies.  
\end{abstract}
\maketitle
Magnetic tunnel junctions (MTJ) based on MgO as the insulator have attracted a lot of attention following the theoretical prediction \cite{slon,berger} and the subsequent experimental confirmation \cite{kise,kat,ralph1,kubota} that a free-layer nano-magnet could be switched using a spin polarized current, commonly referred to as spin-transfer torque (STT). Critical current densities required for switching a typical CoFeB-MgO-CoFeB trilayer structure are usually of the order of $10^{7} A/cm^{2}$ which requires about 1 V of bias for tunnel barriers that are a few nanometers thick \cite{niladri}. The object of this letter is to propose the use of double barrier resonant tunneling as a route to lower the switching voltage, which not only reduces the power dissipation, but also helps to avoid the degradation of the tunnel barrier.\\
\begin{figure}
	\centering
\includegraphics[width=3.4in,height=2in]{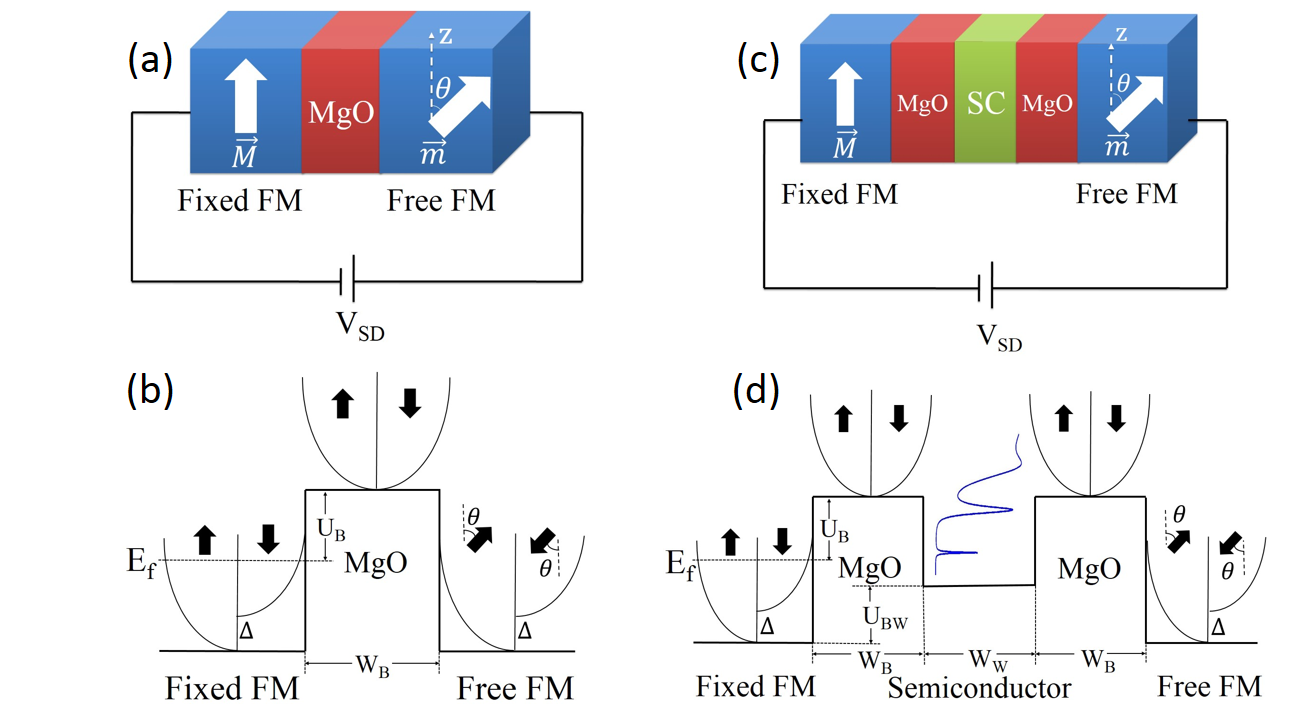}
		\caption{Device details: (a) Schematic of a typical trilayer MTJ. A MgO (Red) layer is sandwiched between two ferromagnetic contacts. The magnetization of the left contact is fixed ($\vec{M}$) along the z-axis, while that of the right contact is free ($\vec{m}$) which makes an angle $\theta$ with the z-axis. (b) Energy band diagram of a trilayer MTJ. The ferromagnetic contacts have an exchange energy of $\Delta$ with $E_f$ being the Fermi energy, $U_B$, the barrier height in the MgO layer above $E_f$. (c) Schematic of the proposed RTMTJ. A semiconductor (green) forms a heterostructure with MgO which is sandwiched between a fixed and free ferromagnetic contact. (d) Energy band diagram of the RTMTJ. $U_{BW}$ is difference in the bottom of the conduction band of the ferro-magnet and the semiconductor. $W_B$ and $W_W$ are the barrier and well widths respectively. Also shown is a sample transmission spectrum of the device with resonant peaks that arise due to confinement.}
		\label{Fig1}
		\end{figure}
\indent 
In this letter, we propose a resonant tunneling magnetic tunnel junction (RTMTJ) structure which results when a heterostructure quantum well is sandwiched between a fixed and a free magnet. This device aims to overcome the fundamental issue presented by a trilayer structure, which is the monotonic opacity to tunneling as a function of barrier width, as depicted in Figs.~\ref{Fig1}(a) and \ref{Fig1}(b). Also, as seen in Figs.~\ref{Fig1}(c) and ~\ref{Fig1}(d), our device offers a larger design landscape with the choice of the semiconductor via the band offset $U_{BW}$ and the variation of the well width $W_W$. We note that earlier proposals involving penta-layer structures \cite{datta1,vedya,theo,salah,niladri}, unlike ours, relied on using an intermediate free magnetic layer for switching. \\
\indent To substantiate our claims, we present RTMTJ designs by employing the non-equilibrium Green's function (NEGF) based spin transport self consistently coupled with the Landau-Lifshitz-Gilbert-Slonczewski equation (LLGS) to describe the free magnet dynamics. As a first demonstration, we show that the physics of resonant tunneling leads to an ultra-high zero-bias tunnel magnetoresistance (TMR) and several non-trivialities in the TMR characteristics with an applied bias. We then present two RTMTJ designs based on MgO-Ge and MgO-ZnO heterostructures and establish that they present a much higher transverse spin torque, thus facilitating lower switching voltages in comparison with trilayer structures. \\
\indent The schematic of the structures studied are shown in Figs.~\ref{Fig1}(a) and ~\ref{Fig1}(c), both of which are connected to a fixed magnet with magnetization $\vec{M}$ taken along the $\hat{z}$ axis, and a free magnet with a magnetization $\vec{m}$ whose thickness is much lesser than that of the fixed layer, thus enabling its switching via spin-transfer torque. The trilayer MTJ has an insulating layer of MgO sandwiched between the two magnetic contacts, while the proposed RTMTJ has a suitable non-magnetic semiconductor that forms a type I or II heterojunction with MgO connected to the two ferromagnetic contacts. Here, we present designs based on Ge and ZnO \cite{petti,shi}, which demonstrate the proof of concept. However, semiconductors such as GaAs\cite{lu}, AlN\cite{yang2} and InN\cite{zhangv}, among others, are also known to form quantum well structures with MgO.\\
\indent All our simulations use CoFeB as ferro-magnets with the equilibrium Fermi level at $E_f=2.25$ eV, and exchange splitting energy $\Delta = 2.15$ eV. The electronic effective mass in either contact is $m^*_{FM}$ = 0.8 $m_e$, and in MgO is $m^*_{ox}$ = 0.18 $m_e$. The CoFeB-MgO interface barrier height is taken to be $U_B$ = 0.76 eV \cite{kubota,datta3}. For the RTMTJ device, the effective masses of Ge and ZnO are $m^*_{Ge} = 0.55 m_e$ and $m^*_{ZnO} = 0.29 m_e$ \cite{sze}, and the well heights are $U_{BW} = -0.51$ eV \cite{petti} and $U_{BW} = -0.23$ eV \cite{shi} respectively. \\
\indent Charge and spin currents are calculated using the non-equilibrium Green's function formalism \cite{datta1} within the effective mass Hamiltonian framework \cite{datta2,butler,lunds,akshay,yanik}. The current operator $I_{op}$ between two lattice points $i$ and $i+1$ is given by $I_{op,i,i+1}=\frac{i}{\hbar}(H_{i,i+1}G^{n}_{i+1,i}-G^{n\dagger}_{i,i+1}H^{\dagger}_{i+1,i})$, following which the charge currents $I$ and the spin currents $\vec{I}_{S}$ are given by
\begin{eqnarray}
I &=& q \int dE \text{ Real [Trace(}I_{op}\text{)]}  \label{ccurrent}  \\
\vec{I}_{S} &=& q \int dE \text{ Real [Trace(}\vec{\sigma}\, I_{op}\text{)]}.
\label{scurrent}
\end{eqnarray}
Here the current operator $I_{op}$ is a 2$\times$2 matrix in spin space, $H$ is the Hamiltonian of the system, $G^n$ is the electron correlation matrix, $\vec{\sigma}$ is the Pauli spin vector, $q$ is the electron charge and $\hbar$ is the reduced Planck constant. For the cross-section, we follow the uncoupled transverse mode approach, with each mode calculated by solving the sub-band eigenvalue problem \cite{kurni,wang} and \eqref{ccurrent} and \eqref{scurrent} representing the charge and spin currents through each mode. In all our simulations, we have summed the currents over the first 50 transverse modes \cite{salah2,niladri}.\\
\indent Spin torque induced magnetization dynamics of the free magnet layer is described by the Landau-Lifshitz-Gilbert-Slonczewski (LLGS) equation\cite{brat,slon}:
\begin{eqnarray}
(1+\alpha^{2})\frac{\partial \hat{m}}{\partial t} &=& -\gamma \hat{m} \times \vec{H}_{eff} - \gamma \alpha (\hat{m} \times ( \hat{m} \times \vec{H}_{eff})) - \vec{\tau}_{S}
\nonumber \\
\vec{\tau}_{S} &=& \frac{\gamma \hbar}{2qM_{S}V}\left[(\hat{m} \times (\hat{m} \times \vec{I}_{S})) - \alpha(\hat{m} \times \vec{I}_{S})\right] \nonumber
\label{taus}
\end{eqnarray}
where $\hat{m}$ is the unit vector along the direction of magnetization of the free magnet, $\gamma$ is the gyro-magnetic ratio of the electron, $\alpha$ is the Gilbert damping parameter, $\vec{H}_{eff} = \vec{H}_{app} + (\frac{2K_{u2}}{M_{S}})m_{z}\hat{z} - (\frac{2K_S}{M_S})m_{x}\hat{x}  $ is the effective magnetic field and $\vec{\tau}_S$ is the spin torque on the free magnet. $\vec{H}_{app}$ is the applied external field, $M_S$ is the saturation magnetization of the free layer, $K_{u2}$ is the uni-axial anisotropy and $2K_{S}/M_{S}$ = 4$\pi M_{S}$ is the demagnetizing field, with $V$ being the volume of the free magnet.\\
\indent In our calculations we resolve the spin current vector \cite{datta3} as $\vec{I}_{S} = I_{S,m}\hat{m} + I_{S,M}\hat{M} + I_{S,\perp}(\hat{M}\times\hat{m})$, where $I_{S,\perp}$ along $\hat{M}\times\hat{m}$ is its transverse component and $I_{S,M}$ along $\hat{M}$ is its parallel component written henceforth as $I_{S,\parallel}$. In the results that follow, the parameters for the magnetization dynamics are $\alpha$ = 0.01, $\gamma$ = 17.6 MHz/Oe, the saturation magnetization $M_{S}$ = 1100 emu/cc and $K_{u2}$ = 2.42 $\times10^{4}$ erg/cc. The cross-sectional area of all devices considered are 70 $\times$ 160 nm\textsuperscript{2} with the thickness of the free magnetic layer taken to be 2 nm.\\
\begin{figure}
	\centering
		\includegraphics[width=3.5in,height=2.6in]{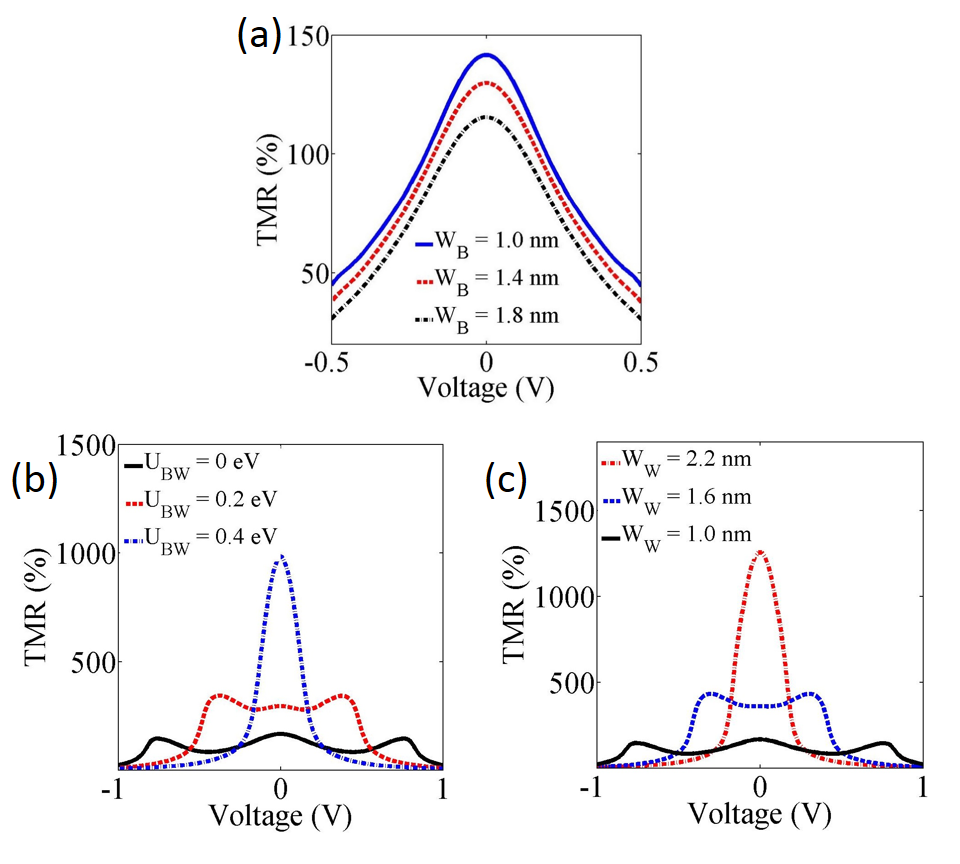}
		\caption{TMR Characteristics (a) for a trilayer MTJ with varying barrier widths $W_B$, (b) for the RTMTJ with barrier width $W_B$ = 1 nm and $W_W$ = 1 nm for different $U_{BW}$ and (c) for the RTMTJ with barrier width $W_B$ = 1 nm and $U_{BW}$ = 0 eV for varying well widths $W_W$. In the case of (a), the characteristic TMR falloff with voltage is noted \cite{datta2}. For (b) and (c), the position of the resonant peaks can be varied to get a sharp rise in the TMR. It is also noted that the position of the resonant peaks can also be controlled using the well width to get a higher TMR.}
\label{Fig2}
\end{figure}
\indent We first explore the TMR characteristics of the two structures with TMR defined as $TMR = (R_{AP}-R_{P})/(R_{P})$, where $R_{P}$ and $R_{AP}$ are the resistances in the parallel (P) and the anti-parallel (AP) configurations respectively. As noted in Fig.~\ref{Fig2} (a), the $W_B$ = 1 nm structure has the highest zero-bias TMR of $140\%$ which drops with increasing barrier width. As reported in literature, our simulations also show a TMR fall-off with bias which can be attributed to the bias dependent potential profile developed across the barrier \cite{datta2}. Our simulations capture the TMR profile for Co$_{0.7}$Fe$_{0.3}$\cite{ralph2} based devices correctly. For different alloys of Co-Fe, it is possible to attain a higher zero-bias TMR ($>$200\%)\cite{parking,yuasag}.\\
\indent For the RTMTJ device, we note in Fig.~\ref{Fig2}(b), that for $U_{BW}$ = 0 eV, the zero-bias TMR is 166\% followed by a semi-oscillatory structure, with a second local peak of 145\%. For $U_{BW}$ = 0.2 eV, one obtains a higher zero-bias TMR of 295\% and a semi-oscillatory behavior with a second peak of 342\%. Finally, for $U_{BW}$ = 0.4 eV, we note a zero-bias TMR of 986\% preceding a sharp fall-off. In Fig.~\ref{Fig2}(c), we present the TMR characteristics for varying well-widths. For $W_W$ = 1.6 nm, the zero bias TMR is 360\% , while for $W_W$ = 2.2 nm, the zero bias TMR reaches 1250\%. It is thus seen that the TMR characteristics of a RTMTJ can be controlled independently by varying the bottom of the well $U_{BW}$, which tantamounts to using different semiconductors, and by altering the well widths.\\
\begin{figure}
	\centering
		\includegraphics[width=3.5in,height=2.5in]{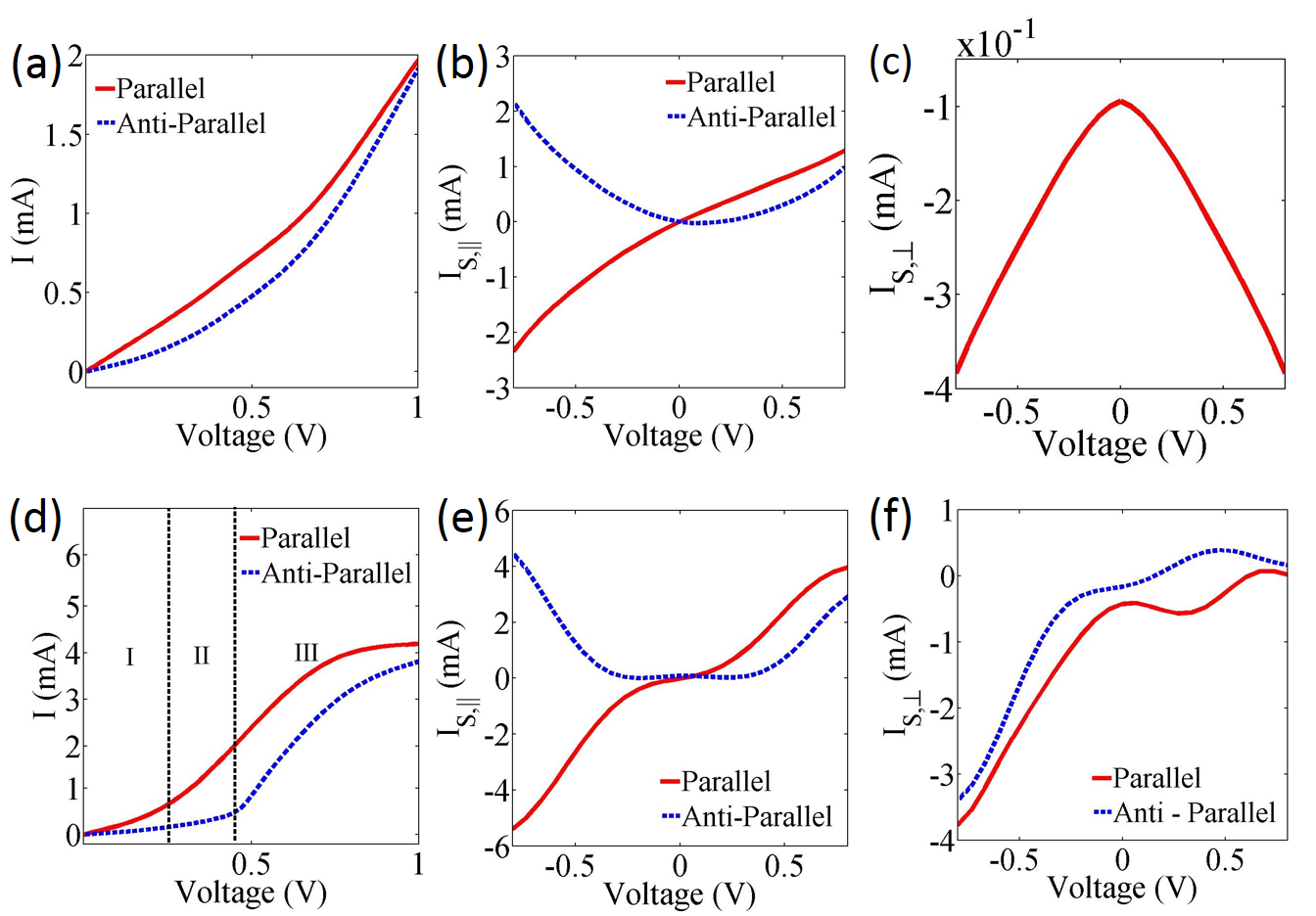}
		\caption{Charge and spin currents of a trilayer MTJ with $W_B$ = 1.6 nm and a RTMTJ with $W_B$ = 1 nm, $W_W$ = 1 nm and $U_{BW}$ = 0.2 eV (a) $I-V$ plots for the trilayer device in the P and AP configurations. (b) $I_{S,\parallel}$-$V$ and (c) $I_{S,\perp}-V$ plots for the trilayer device in the P and AP configurations. The $I_{S,\perp}$ plot here is identical in the P and AP configurations. (d) The $I-V$ plot for the RTMTJ may be divided into three regions (see text) I- tunneling, II- resonant conduction for P state and III-resonant conduction for AP state. (e) $I_{S,\parallel}-V$ and (f) $I_{S,\perp}-V$ plots for the RTMTJ.}
\label{Fig3}
\end{figure}
\indent To understand this exceptional as well as unusual TMR performance of the RTMTJ, we turn to the I-V plots in Fig.~\ref{Fig3}, and also noting that the resonant conduction peaks (see inset of Fig.~\ref{Fig1}(d)) occur at different energies for the P and AP configuration. Fig.~\ref{Fig3}(a) shows the  I-V plots of the trilayer device in the P and the AP state. As expected, the difference between the P and the AP current decreases with voltage leading to a fall in the TMR. The I-V plots for the RTMTJ in Fig.~\ref{Fig3}(d), can however be broadly divided into three regions. In region I, the low current in both the P and AP states is primarily due to tunneling contributions. In region II, the P state current rises due to its first resonant peak falling within the bias window. This leads to the rise in the TMR with voltage as noted in Figs.~\ref{Fig2}(b) and ~\ref{Fig2}(c). In region III, the current in the AP state also starts to increase as its first resonant conduction channel opens up thus explaining the fall in the TMR after its rise in region II. \\
\begin{figure}
	\centering
		\includegraphics[width=3.5in,height=2.6in]{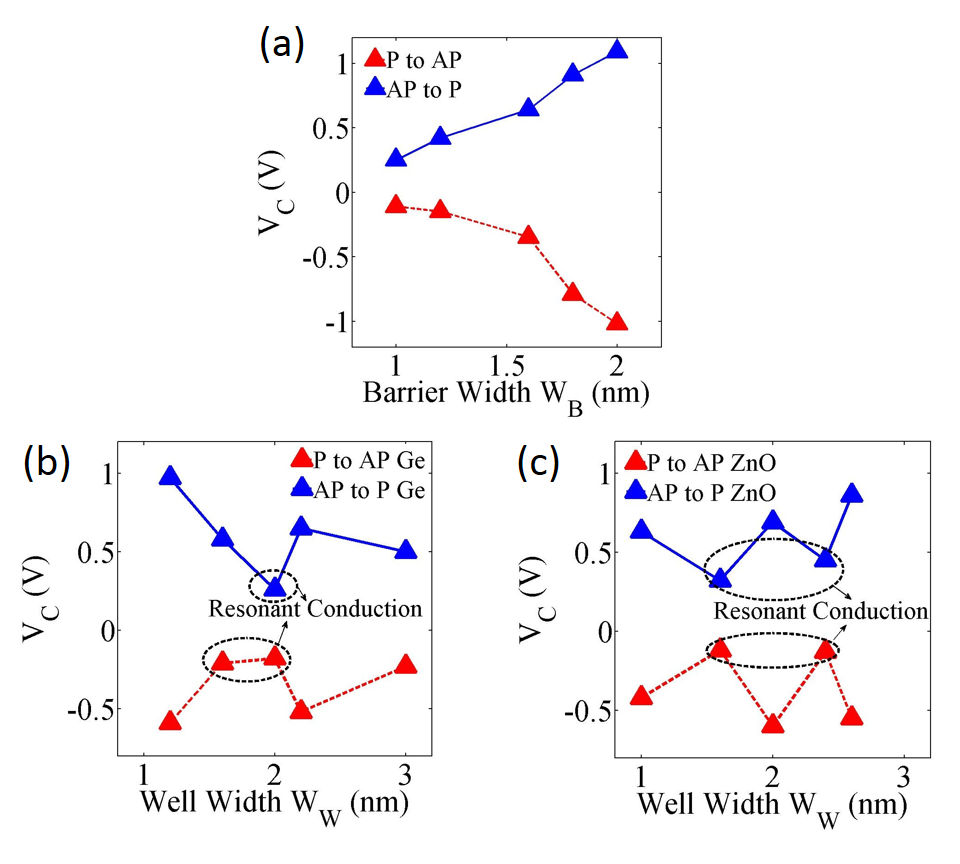}
		\caption{ (a) Critical voltage $V_C$ vs. barrier width $W_B$ for a trilayer device, as the barrier width increases the spin current decreases sharply and a much higher voltage is required for switching. Critical voltage $V_C$ vs. well width $W_W$ for a RTMTJ with (b) Ge and (c) ZnO forming the heterostructure. The width of the barrier in both these cases is $W_{B}$ = 1.2 nm. For Ge we see at $W_W$ = 1.6 nm and 2 nm $V_C$ is very low due to resonant conduction. The same is the case for ZnO at $W_W$ = 1.6 nm and 2.4 nm}
\label{Fig4}		
\end{figure}
\indent In Figs.~\ref{Fig3}(b) and ~\ref{Fig3}(e), we present the parallel spin current $I_{S,\parallel}-V$ plots of a trilayer MTJ and the RTMTJ device respectively. In both the P and AP states, we see that once the resonant state kicks in, the magnitude of the spin current is much higher. By comparing the perpendicular spin current $I_{S,\perp}-V$ plots of the two devices in Fig.~\ref{Fig3}(c) and ~\ref{Fig3}(f), it is seen that while the plots for both the P and AP states are similar in the case of the trilayer device, the plots differ for the P and AP states in the RTMTJ. This is due to the difference in the positions of the resonant peaks for each configuration. It is also observed that the $I_{S,\perp}$ of the RTMTJ is an order of magnitude higher. Thus, if the device parameters of the RTMTJ are chosen such that that the resonant peaks in both the P and AP states are close to the Fermi level of the contact, then an increase in the spin current and torque may be engineered, thus reducing the critical switching voltage $V_C$.\\
\indent Upon examining the variation of $V_C$ of trilayer device with varying barrier widths in Fig.~\ref{Fig4}(a), we note as expected, that the magnitude of $V_{C}$ increases with increasing barrier width. Both the charge and spin currents reduce with increasing barrier width due to the monotonic tunneling opacity of a single tunnel barrier. \\
\begin{figure}
	\centering
		\includegraphics[width=3.5in,height=2.6in]{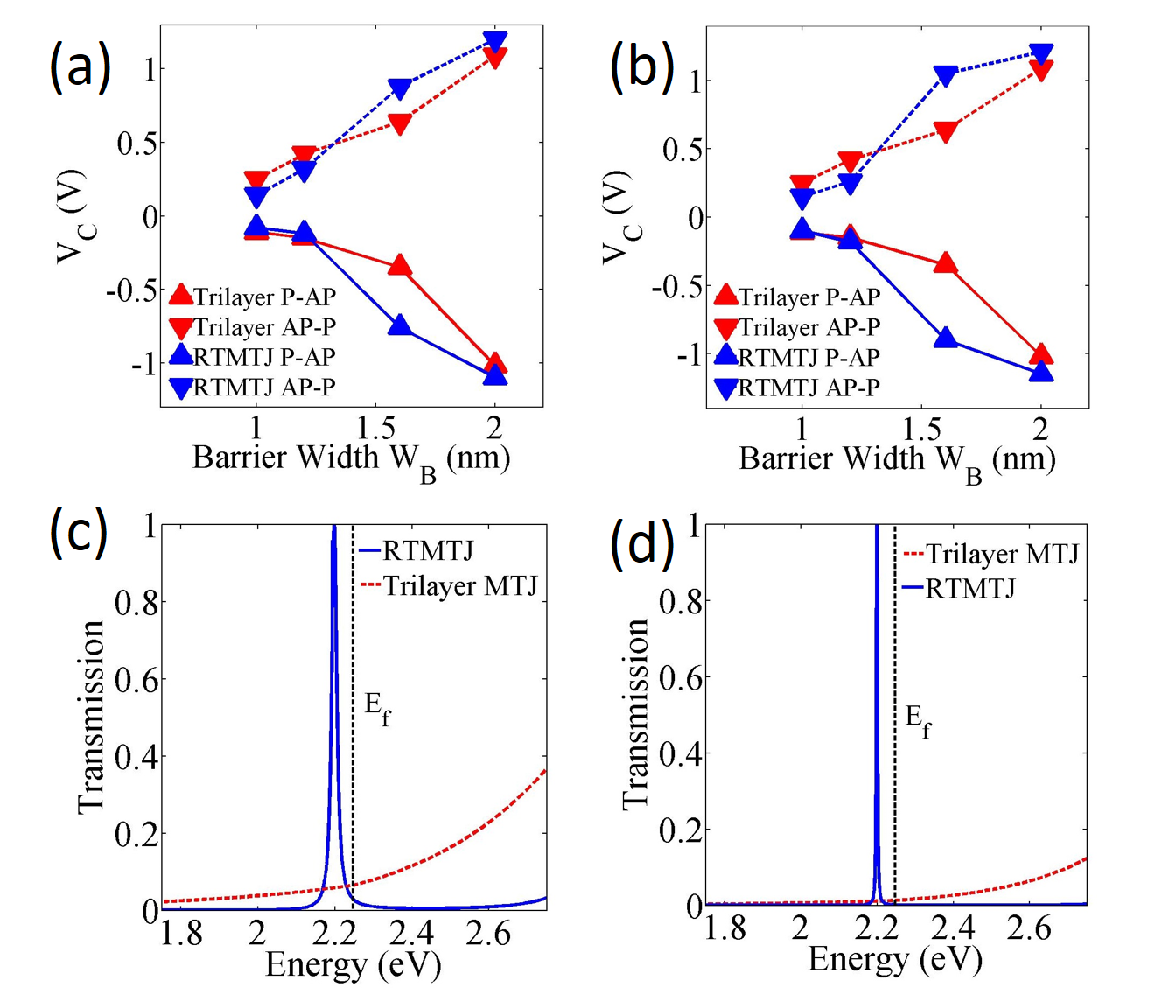}
		\caption{$V_C$ vs. $W_B$ for trilayer MTJs and for well width optimized RTMTJs with (a) ZnO and (b) Ge as the semiconductor. We observe a lower $V_C$ for barrier widths $W_B$ = 1 nm (up to 44\%) and 1.2 nm (up to 38\%) in both cases. Transmission spectra of the lowest order transverse modes of the trilayer MTJs and RTMTJs for (c) $W_B$ = 1.2 nm and (d) $W_B$ = 1.6 nm. We see that the resonant peak is much more broadened for a device with smaller barrier width.}
\label{Fig5}
\end{figure}
\indent Next, we fix the barrier width at $W_{B}$ = 1.2 nm for the RTMTJ structure and demonstrate a reduction in $V_C$. In each case, i.e., Ge and ZnO, the well widths $W_W$ are varied to position the resonant peaks close to the Fermi level of the contacts. In comparing $V_{C}$ of the RTMTJ with that of a trilayer MTJ with $W_B$ = 1.2 nm, it is seen that even though the combined width of the barriers is twice that of the trilayer MTJ, a clear reduction in $V_C$ is noted. For Ge, we observe this at $W_W$ = 2 nm with a switching voltage for P to AP as $V_C$ = -0.18 V (20\% increase) and AP to P as $V_C$ = 0.26 V (38\% lower). For ZnO, we observe a low $V_C$ at both $W_W$ = 1.6 nm and 2.4 nm. For $W_W$ = 1.6 nm, P to AP switching is at $V_C$ = -0.12 V (20\% reduction) and AP to P switching is at $V_C$ = 0.32 V (23\% reduction), while for $W_W$ = 2.4 nm, P to AP switching is at $V_C$ = -0.13 V (13\% reduction) and AP to P switching is at $V_C$ = 0.45 V (7\% increase).\\
\indent Finally, we compare $V_C$ of a trilayer MTJ with that of an RTMTJ for different barrier widths. We have chosen the RTMTJ well widths such that the resonant peak is close to the Fermi level ($E_f$ = 2.25 eV). For both ZnO (Fig.~\ref{Fig5}(a)) and Ge (Fig.~\ref{Fig5}(b)) based RTMTJs, critical switching voltages are lower at smaller barrier widths, i.e., up to 44 \%  lower for $W_B$ = 1 nm and up to 38 \% lower for $W_B$ = 1.2 nm. At larger barrier widths, the trilayer MTJ has a much lower $V_C$ until when $W_B$ = 2 nm, where the critical voltages are almost the same. This curious trend may be understood by analyzing the transmission spectrum. At lower barrier widths, such as at $W_B$ = 1.2 nm, the resonant peak of the RTMTJ is broadened (Fig.~\ref{Fig5}(c)) due to contributions from tunneling and the trilayer transmission around the Fermi level is also significant. The broadened peak of the RTMTJ however leads to a higher current contribution at finite bias, thereby resulting in a lower $V_C$. At larger barrier widths such as at $W_B$ = 1.6 nm,  the resonant peak of the RTMTJ is significantly sharper (Fig.~\ref{Fig5}(d)) and the trilayer device simultaneously suffers a lower transmission near the Fermi level. However, the reduction in the trilayer transmission is less pronounced than the effect of sharpening of the resonant peak in the RTMTJ (Fig.~\ref{Fig5}(d)) which significantly inhibits off-resonant conduction. At $W_B$ = 2 nm, the critical switching voltages for both the trilayer MTJ and RTMTJ are almost the same. At these barrier widths, the trilayer transmission is reduced even further and hence begins to perform as well as the sharpened RTMTJ transmission. Although, it should be noted that apart from the barrier width, the broadening of the resonant peak is also dependent on the semiconductor used. If a suitable semiconductor is chosen such that the resonant peaks are optimally broadened, it should be possible to obtain a low $V_C$ for a wide range of barrier widths.\\
\indent We have thus proposed and explored the use of resonant tunneling as means to enhance the switching characteristics of MTJ devices. In the proposed RTMTJ device, we have demonstrated that the resonant tunneling phenomenon leads to many non-trivial ramifications including an ultra high zero bias TMR and that many interesting TMR regimes may be engineered. Importantly, it is identified that a larger spin-transfer torque under resonance conditions leads to significant reduction in the critical switching voltages (up to 44\%) in comparison with trilayer devices. We believe that the proof-of-concept presented here might lead to practical designs catered towards more efficient STT switching devices and other novel TMR based applications. Specifically, it would be a fruitful venture to investigate similar penta-layer structures that many be engineered to enhance thermal spin torque switching \cite{Bauer_TSTT} via thermoelectric injection enhancement \cite{akshay}. \\\
{\it{Acknowledgments:}} This work was in part supported by the IIT Bombay SEED grant and the Department of Science and Technology (DST), India, under the Science and Engineering Board grant no. SERB/F/3370/2013-2014. The author NC would like to acknowledge useful discussions with Harpreet Arora, Akshay Agarwal and Kantimay Dasgupta. 
\bibliographystyle{apsrev}
\bibliography{refSTT}

\end{document}